\documentclass{article}
\usepackage{amsfonts}

\usepackage{graphicx}
\usepackage{amsmath}

\newtheorem{theorem}{Theorem}

\newtheorem{lemma}[theorem]{Lemma}

\newenvironment{proof}[1][Proof]{\textbf{#1.} }{\ \rule{0.5em}{0.5em}}
\input{tcilatex}

\begin{document}

June 2/03

\bigskip 

\begin{center}
\textbf{Microscopic Space Dimensions and the Discreteness of Time}

\textbf{\bigskip }

\textbf{Abstract}
\end{center}

We present a simple dynamical systems model for the effect of invisible
space dimensions on the visible ones. There are three premises. A: Orbits
consist of flows of probabilities [P].which is the case in the setting of
quantum mechanics. B: The orbits of probabilities are induced by (continuous
time) differential or partial differential equations. C: Observable orbit
are flow of marginal probabilities where the invisible space variables are
averaged out. A theorem is presented which proves that under certain general
conditions the transfer of marginal probabilities cannot be achieved by
continuous time dynamical systems acting on the space of observable
variables but that it can be achieved by discrete time dynamical systems.

\bigskip 

\textbf{1. Introduction}:

The realm of general relativity consists of massive structures and great
distances while the realm of quantum mechanics consists of the smallest
structures such as photons and quarks. In most circumstances one or the
other theory applies without conflict. However, in extreme situations, such
as black holes, both theories are needed for accurate theoretical analysis.
Unfortunately, on very small scales, these two theories are utterly
incompatible. \ 

The assumption of space and time continuity together with the machinery of
calculus are the pillars on which the theories of quantum mechanics and
general relativity rest. Each of these theories has its natural realm;
quantum mechanics (QM), the very small scales, and general realtivity (GR),
the very large. But there are situations where the two must coexist, such as
in black holes. As subatomic particles possess mass and space possesses
structure on all scales, its has been a great challenge for many decades to
apply GR on very small scales where the smooth spatial structures of large
scales give way to precipitous landscapes. The inability to predict dynamics
on these tiny scales is captured by the ineluctable Heisenberg Principle. It
is this fact of nature that forces upon us the reality that point orbits are
meaningless, that the only meaningful dynamic trajectories on these scales
are those of probabilities which, as the scales become larger , are
supported on narrower segments of spacetime, thereby delineating accurately
trajectories of points.

As our objective is to present an idea toward the melding of QM and GR we
shall deal exclusively with orbits of probabilities. We will present an idea
in the form of an example. The idea is to simply untether time from the
constraint of continuity. The motivation for this is based on the fact that
the constraints of continuity on time and space are so great that only very
restricted, scale dependent behavior is possible. If the modern theory of
nonlinear dynamics (chaos) has revealed anything it is that the range of
behavior of even the simplest discrete time systems is incredibly rich and,
as we shall argue, rich enough to accommodate the behavior of particles in
extreme situations where the effects of gravitational attraction on
particles must be taken into account.

In order to capture the larger range of dynamical behavior needed in extreme
situations, string theory purportedly resolves the incompatibility problem
by modifying the equations of general relativity on small scales. But there
is a price for this accomplishment in that - to account accurately for
quantum effects - space is attributed to have, not three, but nine
dimensions. Three dimensions are visible, while the other six dimensions are
curled up in tiny, essentially invisible, strings. The 'extra' dimensions
provide the additional freedom needed to model the dynamics of a unified QM\
and GR\ theory. It is interesting that, in the course of effecting the long
sought unification of QM\ and GR, string theory itself points in the
direction of discrete space and time.

It is the objective of this note to suggest that the breaking of the
shackles of continuous time at the very outset can yield new insights. In
this note we present a simple dynamical system model for the effect of
invisible space dimensions on the visible ones. There are three premises on
which our model is based:

A. Orbits consist of flows of probabilities [P].which is the case in the
setting of quantum mechanics.

B. The orbits of probabilities are induced by (continuous time) differential
or partial differential equations.

C. Observable orbit are flow of marginal probabilities where the invisible
space variables are averaged out.

\bigskip

In section 2 we present the framework and notation of this note. In section
3 we state our main result, which is a dynamcial equivalence between
additional space dimensions and the discreteness of time. In section 4 we
present 2 examples.

\textbf{2. Framework and Notation}

It is now common knowledge that even simple one dimensional maps have the
ability to describe very complicated dynamical behavior of biological and
mechanical systems. Modeling dynamics by a map offers much more variety of
behavior than do differential equations whose solutions are greatly
restricted by time continuity and cannot exhibit chaotic behavior in low
dimensions. Describing dynamical behavior by iterating a map, which can
arise as a Poincare section or by direct modeling offers many benefits from
an analysis perspective. Once a map is determined, the long term statistical
behavior is described by a probability density function, which can be
obtained by measurement of the system or by mathematical means using the
Frobenius-Perron operator \cite{BGbook} as follows: let $I=[0,1]$ denote the
state space of a dynamical system and let $\tau :I\rightarrow I,$ describe
the dynamics of the system. The dynamics is described by a probability
density function $f$ associated with the unique (absolutely continuous
invariant) measure $\mu .$ This is stated mathematically by the following
equation:

\begin{equation*}
\int_{A}fdx=\int_{\tau ^{-1}A}fdx
\end{equation*}
for any (measurable) set \ $A\subset \mathbb{R}.$ \ The Frobenius-Perron
operator, $P_{\tau }f,$ acts on the space of integrable functions and is
defined by

\begin{equation*}
\int_{A}fdx=\int_{A}P_{\tau }fdx
\end{equation*}

The operator $P_{\tau }$ transforms a pdf into a pdf under the
transformation $\tau $. If $\tau $ is piecewise smooth and piecewise
differentiable on a partition of n subintervals, then we have the following
representation for $P_{\tau }$ \cite[Chapter 4]{BGbook}$:$

\begin{equation}
P_{\tau }f(x)=\sum_{z\in \{\tau ^{-1}(x)\}}\frac{f(z)}{\left| \tau ^{\prime
}(z)\right| }
\end{equation}
where for any $x$, the set $\{\tau ^{-1}(x)\}$ consists of at most n points.
The fixed points of \ $P_{\tau }$ are the pdf's.\ 

\textbf{3.Main Result}

Let $\Pr_{n}:R^{n}xR^{m}\rightarrow R^{n}$ be the orthogonal projection.

\begin{theorem}
\ Let\textbf{\ }$X$ \ be a bounded subdomain of $\ R^{n}xR^{m}$ and let $%
\Phi _{t}(x)$ be a continuous dynamical system on $X$ \ (such as generated
by an ordinary or partial differential equation). \ Le $F$ be a compact,
connected subset of \ $X$ and let $f(x)$ be a $C^{1}$ probability density
function supported on $F.$ \ We assume that at time $t=1,$ the dynamical
system $\Phi _{t}(x)$ transfers $f(x)$ to the probability density function $%
g(x)$\ which has support $G$\ and such that $\Pr_{n}(G)\setminus
\Pr_{n}(F)\neq \phi .$\ We assume additionally that both $\Pr_{n}(G)$ and $%
\Pr_{n}(F)$\ are convex. Let $\ f^{\ast }(y)$\ and $g^{\ast }(y)$ be the
marginals of $f(x)$ and $g(x)$\ on $R^{n}.$ Then no continuous time $n$%
-dimensional dynamical system on $\Pr_{n}(G)$ can transfer $f^{\ast }$\ to $%
g^{\ast },$\ but this can be achieved by an $n-$dimensional discrete time
nonlinear dynamical system.
\end{theorem}

\begin{proof}
A continuous time dynamical system on $\Pr_{n}(G)$ which would transfer $%
f^{\ast }$\ to $g^{\ast }$, would induce a homeomorphism $%
h:\Pr_{n}(G)\rightarrow \Pr_{n}(G)$, (onto), such that $h(\Pr_{n}(F))=%
\Pr_{n}(G).$ Since $\Pr_{n}(G)\setminus \Pr_{n}(F)\neq \phi $ this is
impossible. Now, we will construct a nonlinear transformation $%
T:\Pr_{n}(G)\rightarrow \Pr_{n}(G)$, which transfers $f^{\ast }$\ to $%
g^{\ast }.$ We can represent $\Pr_{n}(G)$ as a union of disjoint sets $%
\Pr_{n}(F)\cup F_{1}\cup \cdots \cup F_{k}$ where each of the summands is $%
C^{1}$diffeomorphic to $\Pr_{n}(G)$. Let $h_{0}:\Pr_{n}(F)$ $\rightarrow
\Pr_{n}(G)$ be the diffeomorphism such that $h_{\ast }(f^{\ast }\lambda
)=g^{\ast }\lambda $ (Lemma 1). Let \ $h_{i}:F_{i}\rightarrow \Pr_{n}(G)$, $%
1\leq i\leq k$, be a diffeomorphism. Let $T$ be a map glued from the
diffeomorphisms $h_{i}$, $0\leq i\leq k.$ Obviously $T_{\ast }(f^{\ast
}\lambda )=g^{\ast }\lambda $, or $P_{T}f^{\ast }=g^{\ast }.$
\end{proof}

\begin{lemma}
Let $F$ and $G$ be convex bounded subsets of $R^{n}$. Let $f$ and $g$ be
normalized $C^{1\text{ }}$densities on $F$ and $G$, respectively. Then there
exists a $C^{1}$ diffeomorphism $h:F\rightarrow G$ such that $h_{\ast
}(f\lambda )=g\lambda $, where $\lambda $ is Lebesgue measure on $R^{n}$.
\end{lemma}

\begin{proof}
\bigskip This follows by Lemma 2 of [M].
\end{proof}

Note that if the invisible variables are essential to the dynamical system
then the flow of marginals should not be achievable by any dynamical system
operating in the space $X$ alone.

In the next section we shall consider two dimensional dynamical systems,
that is, a family of two dimensional diffeomorphisms, $\Phi _{t}(x,y),$
which we may consider to be induced by a deterministic system such as a
differential equation or partial differential equation. By a trajectory of $%
\Phi $, we mean a flow in time of probabilities or probability density
functions. We assume we can only observe the $x$ variable and that the $y$
variable acts in an invisible dimension. In terms of the orbit, this means
that we observe only the marginal probabilities in $x$. We then ask: does
there exist a one dimensional dynamical system, that is, a family of one
dimensional homeomorphisms, $\varphi _{t}(x)$, which induces the orbit of
marginal probabilities. We present an example to show that this is in
general impossible. That is, the invisible variable is essential in
describing the observed system and no one dimensional continuous time
dynamical system can account for the observed one dimensional orbit.
However, if we restrict the orbit to a discrete set of times, the example
shows that the discrete time orbit of marginal probabilities can be achieved
by a nonlinear dynamical system. This suggests a duality between what can be
achieved by invisible space dimensions and nonlinear maps acting in discrete
time. The need for discrete time is not surprising as it is an ineluctable
consequence of string theory itself.

\section{ Examples}

1) \ Consider the 2 dimensional differential equation:

\begin{eqnarray}
x^{\prime } &=&y-\frac{1}{2}  \notag \\
y^{\prime } &=&-x+\frac{1}{2}
\end{eqnarray}
on the unit square $S=[0,1]\times \lbrack 0,1]$. This differential equation
induces a dynamical system $\Phi _{t}(x,y)$ which is a rotation around the
point $(\frac{1}{2},\frac{1}{2})$ and is given by

\begin{equation}
\Phi _{t}(x,y)=\left( 
\begin{array}{cc}
\cos t & \sin t \\ 
-\sin t & \cos t
\end{array}
\right) \left( 
\begin{array}{c}
x-\frac{1}{2} \\ 
y-\frac{1}{2}
\end{array}
\right) +(\frac{1}{2},\frac{1}{2}).
\end{equation}
\bigskip Let us now define a probability density function on S as follows: $%
f_{0}(x,y)=2$ if $(x,y)\in \lbrack 0,0.5]\times \lbrack 0,1]$ and $%
f_{0}(x,y)=0$ if $(x,y)\in (0.5,1]\times \lbrack 0,1].$ Let $t=\pi /2$. Then 
$\Phi _{\pi /2}(x,y)$ transforms $f_{0}(x,y)$ to $f_{1}(x,y)=2$ if $(x,y)\in
\lbrack 0,1]\times \lbrack 0.5,1]$ and $f_{1}(x,y)=0$ if $(x,y)\in \lbrack
0,1]\times \lbrack 0,0.5].$ \ Let us now consider the associated marginal
probability density functions:

\begin{equation*}
g_{0}(x)=\int_{0}^{1}f_{0}(x,y)dy
\end{equation*}

\begin{equation*}
g_{1}(x)=\int_{0}^{1}f_{1}(x,y)dy
\end{equation*}

Then $g_{0}(x)=2$ if $x\in \lbrack 0,.5]$ and $\ g_{0}=0$ if $x\in (.5,1].$
Also, $g_{1}(x)=1$ if $x\in \lbrack 0,1].$ Now there is no homeomorphism
that transforms $g_{0}$ to $g_{1},$that is, there is no one-dimensional
continuous time dynamical system that can transform $g_{0}$ to $g_{1}.$
However, the triangle map can do it in one iteration as we now show. Let $%
T:[0,1]\rightarrow \lbrack 0,1]$ be defined by $T(x)=1-2\left| x-1/2\right|
. $ \bigskip Its Frobenius-Perron Operator is given by

\begin{equation}
P_{T}f(x)=\frac{1}{2}f(\frac{x}{2})+\frac{1}{2}f(\frac{2-x}{2})
\end{equation}

It is easy to see that $P_{T}g_{0}(x)=g_{1}(x).$

\textbf{2) Long thin rod example\bigskip }.

Let \ $A>0$ be a fixed number. We consider the rectangular $S=[0,A]\times
\lbrack 0,1/A]$ as a thin long rod.\ Consider the 2 dimensional differential
equation:

\begin{eqnarray}
x^{\prime } &=&A^{2}y-\frac{1}{2}A  \notag \\
y^{\prime } &=&-A^{-2}x+\frac{1}{2A}
\end{eqnarray}
on the rectangular $S$. This differential equation induces a dynamical
system $\Phi _{t}(x,y)$ which is a distorted rotation around the center of $%
S $ and is given by

\begin{equation}
\Phi _{t}(x,y)=\left( 
\begin{array}{cc}
\cos t & \sin t \\ 
\frac{-\sin t}{A^{2}} & \frac{\cos t}{A^{2}}
\end{array}
\right) \left( 
\begin{array}{c}
x-\frac{1}{2}A \\ 
y-\frac{1}{2A}
\end{array}
\right) +(\frac{1}{2}A,\frac{1}{2A}).
\end{equation}
\bigskip Let us again define a probability density function on $S$ as
follows: $f_{0}(x,y)=2$ if $(x,y)\in \lbrack 0,A/2]\times \lbrack 0,1/A]$
and $f_{0}(x,y)=0$ if $(x,y)\in (A/2,A]\times \lbrack 0,1/A].$ Let $t=\pi /2$%
. Then $\Phi _{\pi /2}(x,y)$ transforms $f_{0}(x,y)$ to $f_{1}(x,y)=2$ if $%
(x,y)\in \lbrack 0,A]\times \lbrack 0.5/A,1/A]$ and $f_{1}(x,y)=0$ if $%
(x,y)\in \lbrack 0,A]\times \lbrack 0,0.5/A].$ \ Let us now consider the
associated marginal probability density functions:

\begin{equation*}
g_{0}(x)=\int_{0}^{A}f_{0}(x,y)dy
\end{equation*}

\begin{equation*}
g_{1}(x)=\int_{0}^{A}f_{1}(x,y)dy
\end{equation*}

Then $g_{0}(x)=2/A$ if $x\in \lbrack 0,A/2]$ and $\ g_{0}(x)=0$ if $x\in
(A/2,A].$ Also, $g_{1}(x)=1/A$ if $x\in \lbrack 0,A].$

Now there is no homeomorphism of $[0,A]$ onto itself that transforms $g_{0}$
to $g_{1},$that is, there is no one-dimensional continuous time dynamical
system that can transform $g_{0}$ to $g_{1}.$ However, an analog of the
triangle map can do it in one iteration as we now show. Let $%
T:[0,A]\rightarrow \lbrack 0,A]$ be defined by

\begin{equation*}
T(x)=\left\{ 
\begin{array}{cc}
2x, & 0\leq x\leq A/2 \\ 
2A-2x & A/2<x\leq A
\end{array}
\right. .
\end{equation*}
\bigskip Its Frobenius-Perron Operator is given by

\begin{equation}
P_{T}f(x)=\frac{1}{2}f(\frac{x}{2})+\frac{1}{2}f(A-\frac{x}{2})
\end{equation}

It is easy to see that $P_{T}g_{0}(x)=g_{1}(x).$

\bigskip

\bigskip \textbf{Summary:}

There is no continuous dynamical system in observable variables that can
account for behavior. However,we can find a discrete time chaotic dynamical
system in the observable dimensions that completely accounts for it.

\bigskip

\textbf{References}

[BG] The Laws of Chaos, A. Boyarsky and P. Gora, Birkhauser, 1997.

[G] The Elegant Universe, B. Green..

[K] Functional Equations in a Single Variable, M. Kuczma, Polish Scientific
Publishers, Warsaw, 1968.

[NS] Qualitative Theory of Differential Equations, V.V. Nemytskii and V.V.
Stepanov, Princeton University Press, 1960.

[P] The End of Certainty, Ilya Prigogine, The Free Press, N.Y., 1996

[L] Can time be a disrete dynamical variable? Physics Letters Vol. 122B, No.
3,4. March 10, 1983.

[M] On the Volume Elements on a Manifold, Jurgen Moser, Trans. Amer. Math.
Soc. 120 (1965), 286--294.

\begin{quote}
\bigskip 

\bigskip 

\bigskip \_\_\_\_\_\_\_\_\_\_\_\_\_\_\_\_\_\_\_\_\_\_\_\_
\end{quote}

In [P] Prigogine argues against the notion of deterministic point
trajectories in dynamical systems, proposing that trajectories of ensembles
are better models of reality \ His arguments are in part due to the fact
that it is impossible to specify points with complete accuracy. \ This is of
course in full accord with quantum mechanics where the object of interest is
the probability amplitude as it is derived from the wave function. This is
also very much the case for chaotic dynamical systems due to the instability
of orbits leading to a statistical description of dynamics [BG].

\end{document}